\documentclass[12pt]{article}

\usepackage{amsmath,amssymb}

\makeatletter

\@addtoreset{equation}{section}
\makeatother

\begin{document}
\baselineskip=16pt
\begin{titlepage}
\begin{flushright}
{\small KYUSHU-HET-70}\\[-1mm]
hep-th/0311233
\end{flushright}
\begin{center}
\vspace*{1.2cm}

{\large\bf%
Anomalies and Fayet-Iliopoulos terms on warped orbifolds\\[1mm] and
large hierarchies}\vspace{1cm}

Takayuki {\sc Hirayama}\footnote{\tt 
hirayama@physics.utoronto.ca} and 
Koichi {\sc Yoshioka}\footnote{\tt
yoshioka@higgs.phys.kyushu-u.ac.jp}\vspace{3mm}

{\it $^*$Department of Physics, University of Toronto,
Toronto, Ontario, M5S1A7, Canada}\\
{\it $^\dag$Department of Physics, Kyushu University,
Fukuoka 812-8581, Japan}\vspace{10mm}

\end{center}

\begin{abstract}\noindent
The anomalies in five-dimensional orbifold theories are examined in a
generic type of non-factorizable geometries. In spite of complicated
fermion wavefunctions, the shape of anomaly is found to be identical 
to that of flat theories. In particular it is split evenly on the
orbifold fixed points. This result also follows from the arguments on
the AdS/CFT correspondence and an anomaly cancellation mechanism. The
cancellation with Chern-Simons term works if the four-dimensional
effective theory is free from chiral anomalies. We also discuss the
Fayet-Iliopoulos (FI) term in warped supersymmetric theories. Unlike
the gauge anomaly, FI divergences reside not only on the orbifold
fixed points but also in the whole five-dimensional bulk. The effect
of the FI term is to generate supersymmetric masses for charged
hypermultiplets, which are no longer constant but have metric factor
dependence. We calculate the spectrum and wavefunctions of
Kaluza-Klein modes in the presence of the FI term and discuss
phenomenological implications to quark-lepton masses and large scale
hierarchies.
\end{abstract}

\end{titlepage}
\setcounter{footnote}{0}

\section{Introduction}

In these years, much progress has been made with theories in higher
dimensions. The geometries of extra dimensional spacetime have been
playing an essential role for explaining various unresolved problems
in particle physics, e.g.\ the hierarchy problem. Among them, a model
which realizes a large mass hierarchy is presented~\cite{RS} by use of
a non-factorizable form of five dimensional metric.

In the model of Ref.~\cite{RS}, all the standard model fields are
confined on a brane. The possibilities that some of matter and gauge
fields propagate in the bulk space has also been pursued
in~\cite{RSext}. There orbifold projections are imposed on the fields
at the boundaries in order to realize chiral spectrum in effective
four-dimensional theories. An orbifold theory is often chiral and may
become an anomalous gauge theory. To be a consistent theory, such
anomaly should be canceled. Although a higher-dimensional field theory
is apparently not renormalizable and a cutoff scale is not so high
compared with the compactification scale, one can discuss quantities
of higher-dimensional theories, such as anomalies and renormalization
of couplings up to the cutoff scale. The gauge anomaly in an orbifold
field theory has been calculated for the flat extra dimensions and
found that the anomaly is localized on the orbifold fixed
points~\cite{ACG,SSSZ}. Even when the low-energy effective theory has
vector-like mass spectrum, gauge anomalies are induced by one-loop
fermion diagrams in higher dimensions. In this case, however, the
total anomaly is canceled out by adding a Chern-Simons term to the
action~\cite{CSterm,BCCRS}. Keeping in mind phenomenological interests 
of non-factorizable geometries, in the first half of this paper we
study gauge anomalies of orbifold theories in a generic type of
non-factorizable geometries (including the metric of~\cite{RS}). We
will find that the anomaly is localized in the exactly same way as in
flat theories and is canceled if the effective four-dimensional theory
is vector like. It might be difficult to understand this evenly-split
anomaly because wavefunctions of fermions are highly complicated
depending on five-dimensional metrics. We will show that the anomaly
should be split equally on the fixed points from the theoretical
arguments on the AdS/CFT correspondence and an anomaly cancellation
mechanism.

\smallskip

When one considers a supersymmetric $U(1)$ gauge theory, the
Fayet-Iliopoulos (FI) term is important for studying the vacuum and
low-energy spectrum of the system. The appearance of FI term is deeply
connected to gravitational anomalies in supergravity theories. For
flat supersymmetric orbifold theories, it is known that quantum
effects induce non-vanishing FI terms on the fixed
points~\cite{exFID,SSSZ} and hence a condition for supersymmetric
vacuum is modified~\cite{AGW,BCCRS}. In the second half of this paper,
we investigate the structure of FI term in supersymmetric
five-dimensional theory on the warped background. Unlike the flat
background, we find that FI divergences are induced not only on the
fixed points but also in the five-dimensional bulk. The
five-dimensional theory has a supersymmetric vacuum in the presence of
bulk and/or boundary fields. In this vacuum FI terms induce bulk
masses of $U(1)$ charged hypermultiplets which depend on the warped
metric factor and are proportional to the coefficients of FI terms. We
discuss the Kaluza-Klein (KK) mass spectrum in the presence of such
bulk mass terms and study characteristic behavior of wavefunction
profiles. In particular, the localization of light modes in the extra
dimension is important for four-dimensional phenomenology. As typical
examples, several realistic models are constructed in this framework
to realize Yukawa and Planck/weak mass hierarchies.

\section{Gauge anomaly in curved spacetime}

For later comparison, we first review the gauge anomaly in flat
five-dimensional orbifold theories. The fifth dimension parameterized
by $y$ is compactified on the $S^1/Z_2$ orbifold with two fixed 
points $y=0$ and $\pi R$. The appropriate boundary conditions at the
fixed points are introduced for a bulk gauge field $A_M(x,y)$ and a
Dirac spinor $\Psi(x,y)$. They are defined so as to be consistent with
the $Z_2$ orbifold:
\begin{eqnarray}
  A_\mu(x,y) &=& +A_\mu(x,-y), \qquad 
  A_\mu(x,y-\pi R) \,=\, +A_\mu(x,-y+2\pi R), \nonumber \\
  A_5(x,y) &=& -A_5(x,-y), \qquad 
  A_5(x,y-\pi R) \,=\, -A_5(x,-y+2\pi R), \\
  \Psi(x,y) &=& +\gamma_5 \Psi(x,-y), \qquad
  \Psi(x,y-\pi R) \,=\, +\gamma_5 \Psi(x,-y+2\pi R). \nonumber 
\end{eqnarray}
The indices of Roman letters $M,\,N,\cdots,$ run over all dimensions
and the Greek letter indices $\mu,\,\nu,\cdots,$ run only over the
first four dimensions. The $4\times 4$ matrix $\gamma_5$ is
diag$\,(-1,-1,+1,+1)$. With these boundary conditions, $A_\mu$ and the
right-handed component $\big(\frac{1+\gamma_5}{2}\big)\Psi$ have
four-dimensional massless modes. Due to the fermion boundary conditions, 
the theory is chiral on the fixed points and the gauge anomaly might
be present in the divergence of gauge 
current $J^M=\bar\Psi\Gamma^M\Psi$. Since the anomaly is a low-energy
effect, its form can be calculated via Kaluza-Klein reduction
procedure and is found~\cite{ACG}
\begin{eqnarray}
  \partial_MJ^M &=& \frac{g^2}{32\pi^2}\epsilon^{\mu\nu\rho\sigma}
  F_{\mu\nu}F_{\rho\sigma} \big[\delta(y)+\delta(y-\pi R)\big].
  \label{flat-anomaly}
\end{eqnarray}
The delta function is defined 
as $\int_0^{\pi R}\!f(y)\delta(y)dy=f(0)/2$. The anomaly 
coefficient $g^2/32\pi^2$ is proportional to the chiral anomaly for a
four-dimensional Weyl fermion. In the effective theory viewpoint, the
anomaly (\ref{flat-anomaly}) comes from the zero-mode contribution and
a Chern-Simons term which is generated by integration of heavy KK
fermions.

Several interesting properties are in the form of gauge 
anomaly (\ref{flat-anomaly}); (i) there is no anomaly in the bulk. It
is localized at the fixed points as expected from the chiral boundary
conditions, (ii) the anomaly is equally split between the two fixed
points, and (iii) the integration of the anomaly over $y$ direction is
nonzero which is understood from the fact that there is a chiral zero
mode. For different boundary conditions of fermions that do not leave
any chiral zero modes, the integrated anomalies in effective theories
vanish~\cite{SSSZ,BCCRS}. Various other aspects of localized anomalies
in flat spacetime have been studied~\cite{others}.

\medskip

Now let us consider five-dimensional gauge theories in curved
spacetime and present the possible form of one-loop gauge
anomalies. Throughout this paper, the gravity is treated as a
background as we are interested in consistency of field theory. We
particularly focus on the following non-factorizable background metric:
\begin{eqnarray}
  ds^2 &=& a^2(y)\eta_{\mu\nu}dx^{\mu}dx^{\nu}+dy^2,
  \label{warp}
\end{eqnarray}
where $\eta_{\mu\nu}$ is the four-dimensional flat Minkowski metric,
and we normalize the factor $a(y)$ so that $a(0)=1$. A
phenomenologically interesting example is the anti 
de-Sitter (AdS) geometry where $a(y)$ is given by the warp 
factor $e^{-k|y|}$ ($k$ is the AdS curvature). In this section,
however, we do not specify the metric factor, and will see that the
gauge anomaly on the background (\ref{warp}) does not depend on the
form of $a(y)$.

We consider the action for a gauge field $A_M(x,y)$ and a Dirac 
spinor $\Psi(x,y)$ on the background (\ref{warp})
\begin{eqnarray}
  S &=& \int\!d^4xdy \sqrt{-g} \bigg[-\frac{1}{4}F_{MN}F^{MN}
  +\bar\Psi i\Gamma^MD_M\Psi -m(y)\bar\Psi\Psi \bigg],
\end{eqnarray}
where $\Gamma_M$ are the gamma matrices in five dimensions 
and $D_M$ is the covariant derivative including the spin connection
and the gauge field; $D_M=\partial_M -igA_M+\frac{1}{8}\omega_M^{PQ} 
[\Gamma_P,\Gamma_Q]$. The bulk mass parameter $m(y)$ must obey the
condition $m(-y)=-m(y)$ consistent with the $Z_2$ orbifolding. We will
consider the Abelian example and the extension to non-Abelian gauge
theory is straightforward. The action is classically invariant under
the gauge transformation with a gauge parameter $\Lambda(x,y)$
\begin{eqnarray}
  A'_M(x,y) &=& A_M(x,y) +\partial_M \Lambda(x,y), \\
  \Psi'(x,y) &=& e^{i\Lambda(x,y)} \Psi(x,y).
\end{eqnarray}
The corresponding conserved current $(\eta^{MN}\partial_MJ_M=0)$ is 
\begin{eqnarray}
  J_M &=& \sqrt{-g}\bar{\Psi}\Gamma_M\Psi \;=\; 
  \left\{\begin{array}{lccl}
      a^3(y)\bar{\Psi}\gamma_\mu\Psi &&& M=\mu \\[2mm]
      a^4(y)\bar{\Psi}(i\gamma_5)\Psi &&& M=5 \ , \\
    \end{array}\right. 
  \label{current}
\end{eqnarray}
where $\gamma_\mu$ are the gamma matrices in four dimensions.

As in the flat case, we compactify the fifth dimension on a line
segment $y=[0,\,\pi R]$ and introduce a brane at each
boundary.\footnote{The finite four-dimensional Newton constant can be
realized with infinite extra dimension(s) depending on the factor
$a(y)$. Note, however, that the zero modes of gauge fields have
constant wavefunctions along $y$ direction and the normalizability
must require a finite size of extra dimension (two boundaries) in the
absence of some localization mechanism for the zero modes.}
The orbifold boundary conditions for bulk fields 
at $y=0$ and $y=\pi R$ are also similar to the flat case. In
particular, the Dirac fermion is subject to the following parity
conditions:
\begin{eqnarray}
  \Psi(x,y) \,=\, \gamma_5 \Psi(x,-y), \qquad
  \Psi(x,y-\pi R) \,=\, \eta \gamma_5 \Psi(x,-(y-\pi R)).
  \label{bc-warp}
\end{eqnarray}
For the conditions consistent with the $Z_2$ orbifold, $\eta$ has to
satisfy $\eta^2=1$. For $\eta=+1$, we have a massless right-handed
chiral fermion in the low-energy effective theory. On the other
hand, for $\eta=-1$, all KK fermions become massive, and massless
modes are projected out. A five-dimensional gauge theory is expected
to have no chiral anomaly because of its vector-like nature. However,
with the chiral boundary conditions (\ref{bc-warp}), the gauge 
current (\ref{current}) suffers from chiral anomalies at quantum
level. In particular, although the effective theory is vector-like for
the $\eta=-1$ case, we will see the anomalies are induced onto the
boundaries.

Now we calculate the chiral anomaly on the curved background 
metric (\ref{warp}) via the KK point of view. We decompose $\Psi$ into
the four-dimensional KK modes
\begin{eqnarray}
  \Psi(x,y) &=& \sum \psi^+_n(x)\chi^+_n(y) +\psi^-_n(x)\chi^-_n(y),
\end{eqnarray}
where $\psi_n^{+ (-)}(x)$ are four-dimensional right-(left-) handed
fermions. The equations of motion are given by
\begin{eqnarray}
  \Big[\partial_y +2\frac{\partial_y a}{a}-m(y)\Big]\chi^+_n &=& 
  +\frac{m_n}{a}\chi^-_n, \\
  \Big[\partial_y +2\frac{\partial_y a}{a}+m(y)\Big]\chi^-_n &=& 
  -\frac{m_n}{a}\chi^+_n,
\end{eqnarray}
where $m_n$ are the four-dimensional masses ($p_\mu^2=-m_n^2$). In
what follows, we consider a massless Dirac fermion $m(y)=0$ for an
illustrative example. However, it will be found that the shape of gauge
anomalies is insensitive to a value of $m(y)$ and therefore the
analysis with $m(y)=0$ gives generic results. The mass eigenmodes 
for $m(y)=0$ are
\begin{eqnarray}
  \chi^+_n(y) &=& \sqrt{\frac{2}{A(\pi R)}}\,a^{-2}(y) 
  \cos\big[m_nA(y)\big], \label{chip} \\
  \chi^-_n(y) &=& \sqrt{\frac{2}{A(\pi R)}}\,a^{-2}(y) 
  \sin\big[m_nA(y)\big], \label{chim} 
\end{eqnarray}
and $A(y)\equiv\int^y_0\!dy'a^{-1}(y')$. These eigenfunctions satisfy
the orbifold conditions at the $y=0$ boundary. The normalization
factors are determined so that the four-dimensional modes have
canonical kinetic 
terms $\int_0^{\pi R}\!dy\,a^3(y)\chi^\pm_m(y)\chi^\pm_n(y)\,
=\,\delta_{mn}$. The mass eigenvalues $m_n$ are fixed by the boundary
condition at $y=\pi R$,
\begin{eqnarray}
  m_n &=& \left\{
    \begin{array}{ccl}
      \displaystyle{\frac{n\pi}{A(\pi R)}} && (\eta=+1) \\[3mm]
      \displaystyle{\frac{(n+1/2)\pi}{A(\pi R)}} && (\eta=-1)
    \end{array} \right. \qquad n\,=\,0,1,2,\cdots.
\end{eqnarray}

Along the line of Ref.~\cite{ACG}, we calculate the five-dimensional
gauge anomaly by summing up familiar four-dimensional anomalies of the
KK modes. With noting the metric factor appears in the current, the
form of anomaly is expressed in terms of the eigenfunctions
\begin{eqnarray}
  \eta^{MN}\partial_MJ_M &=& \frac{g^2}{32\pi^2}
  \epsilon^{\mu\nu\rho\sigma}F_{\mu\nu}F_{\rho\sigma}\,a^3(y)
  \Big(\sum |\chi^+_n|^2-\sum |\chi^-_n|^2\Big).
\end{eqnarray}
Substituting the eigenfunctions (\ref{chip}) and (\ref{chim}), the
summation over all the KK modes becomes
\begin{eqnarray}
  \sum |\chi^+_n|^2-\sum |\chi^-_n|^2 &=& \frac{2a^{-4}(y)}{A(\pi R)}
  \Big[\sum \cos^2\big[m_nA(y)\big] -\sum \sin^2\big[m_nA(y)\big]\Big]
  \nonumber \\ 
  &=& \frac{a^{-4}(y)}{A(\pi R)} 
  \bigg[\delta\bigg(\frac{A(y)}{A(\pi R)}\bigg)
  +\eta\delta\bigg(\frac{A(y)}{A(\pi R)}-1\bigg)\bigg] \nonumber \\[2mm] 
  &=& \delta(y)+\eta a^{-3}(\pi R)\delta(y-\pi R).
\end{eqnarray}
We thus obtain the five-dimensional gauge anomaly on the curved 
metric (\ref{warp}):
\begin{eqnarray}
  \eta^{MN}\partial_MJ_M 
  &=& \frac{g^2}{32\pi^2}\epsilon^{\mu\nu\rho\sigma}
  F_{\mu\nu}F_{\rho\sigma} \big[\delta(y)+\eta\delta(y-\pi R)\big].
  \label{warp-anomaly}
\end{eqnarray}From 
this form 
one can see that the anomaly appears only at the boundaries and
furthermore the sizes of anomalies on the two boundaries are
equal. It is interesting that the result is exactly same as that in
the flat spacetime. There is no dependence on the metric 
factor $a(y)$. For a nonzero Dirac mass $m(y)$, the wavefunctions of
fermion KK modes have completely different forms. We have, however,
checked that even in that case, the gauge anomaly is not affected by
the presence of mass term and therefore is given 
by (\ref{warp-anomaly}). If there is a chiral four-dimensional fermion
living on a brane, it induces a localized anomaly with the 
coefficient $g^2/16\pi^2$ at $y=0$ or $y=\pi R$, depending on the
position where the fermion lives.

\medskip

Intuitively, it might be difficult to understand the fact that the
anomaly is split evenly between the two boundaries. This is mainly
because, on the curved metric, the wavefunctions of KK fermions are 
highly curved and generally take rather different values on the two
boundaries. For example, for a fermion in the AdS$_5$ geometry with a
bulk mass smaller than $k/2$, all KK fermions including the
massless mode are strongly peaked at the $y=\pi R$ brane with the
exponential warp factor. Nevertheless our 
result (\ref{warp-anomaly}) shows that the gauge anomaly is not
localized in a lopsided way on that brane. In the following, we argue
two theoretical grounds why it should be so in a five-dimensional
orbifold theory.

The first argument is related to a condition of anomaly
cancellation. As mentioned before, for $\eta=-1$ (without boundary
fermions), the low-energy effective theory has vector-like mass
spectrum and therefore is supposed not to suffer from any
anomalies. This implies that the apparent 
anomaly (\ref{warp-anomaly}) does not introduce quantum breaking of
gauge symmetry but, with appropriate regularization, it should be
canceled by additional local terms which are allowed by the
symmetry. In the present five-dimensional theory, an adequate term for
this purpose is the Chern-Simons term:
\begin{eqnarray}
  S_{\rm CS} &=& \int\!d^4xdy\,\rho(y)\epsilon^{MNPQR}A_{M}F_{NP}F_{QR}.
\end{eqnarray}
The coefficient $\rho(y)$ is periodic and must be an odd function 
of $y$ such that it respects the $Z_2$ symmetry. It is easily found
that the gauge variation of $S_{\rm CS}$ gives rise to the right term
to cancel the anomaly. The vanishing anomaly is ensured if the condition
\begin{eqnarray}
  \frac{g^2}{32\pi^2}\big[\delta(y)-\delta(y-\pi R)\big]-
  \partial_y \rho(y) &=& 0
\end{eqnarray}
is satisfied. We find the solution which is given by
\begin{eqnarray}
  \rho(y) &=& \frac{g^2}{64\pi^2}\epsilon(y),
  \label{coeff}
\end{eqnarray}
where $\epsilon(y)$ is the sign function with 
anti-periodicity $\pi R$ and is allowed by the symmetry. It should be
noticed that the localized anomaly can be canceled by the Chern-Simons
term only if the two coefficients of delta functions in the anomaly
are equal to each other and have a relative minus sign
($\eta=-1$). Thus, this argument suggests that the localized anomaly
should have the form (\ref{warp-anomaly}) and be independent of the
mass parameter $m(y)$. With an appropriate regularization scheme, the
Chern-Simons term with the coefficient (\ref{coeff}) is indeed
generated at loop level. It is also clear that the anomaly 
for $\eta=+1$ cannot be canceled by Chern-Simons terms. This is
because massless modes necessarily appear in the low-energy theory and
the regularization \`a la Pauli-Villars does not work.

\smallskip

Another theoretical support comes from the AdS/CFT
correspondence~\cite{ADSCFT,ADSCFT-anom}, which claims that a
five-dimensional gravity theory on the AdS$_5$ is holographic dual to
a four-dimensional conformal field theory (CFT) in the 
large $N$ limit. It is also argued that an AdS$_5$ theory with
boundaries, e.g.\ the Randall-Sundrum model~\cite{RS}, has a certain
four-dimensional dual description~\cite{ADSCFT-RS, CFT-anom}. The
analysis of four-dimensional dual theory gives an understanding of the
evenly localized anomaly in the AdS$_5$.

In the AdS/CFT correspondence, the fifth dimension is encoded as the
energy scale in the four-dimensional CFT\@. A larger (smaller) value
of the fifth dimension $y$ corresponds to the infrared (ultraviolet)
in four dimensions. Modifying the AdS side by introducing the Planck
brane is interpreted in the CFT as introducing a ultraviolet (UV)
cutoff of the Planck scale and turning on the coupling to gravity. The
structure of Planck brane theory determines the details of UV
deformation of CFT\@. On the other hand, the presence of the IR brane
implies the breaking of conformal symmetry in the IR regime. We are
now interested in a bulk gauge theory $G$ on the AdS background, which
means that the subgroup $G$ of global symmetry in the CFT is weakly
gauged. It is indeed checked that the anomaly in CFT is related to the 
Chern-Simons term in the AdS side~\cite{ADSCFT-anom}.

To examine the anomaly and its cancellation in the CFT side, we need
to know how various types of AdS fields are mapped into the
four-dimensional description under the duality. It is known that
localized  four-dimensional fields on the UV brane are interpreted as
sources of CFT\@. They do not have direct couplings to CFT in the
four-dimensional gravity point of view, but $G$ gauge interactions
connect the two sectors ($G$ is weakly gauged). On the other hand,
four-dimensional fields on the IR brane correspond to massless
composites of the CFT\@. This is suggested from the fact that the IR
brane fields are strongly coupled to the KK-excited gravitons which
are localized on the IR brane. In a similar manner, KK-excited modes
of bulk fields correspond to massive bound states of the broken CFT\@.

Matching of AdS bulk fields is also influenced by boundary conditions
at the two branes. There are four types of boundary conditions on
fermion fields, $\psi^{(\pm\pm)}$. The first and second indices
indicate the boundary conditions at the UV and IR branes,
respectively. The $+$ sign denotes a Neumann type and $-$ a Dirichlet
one. With the notation (\ref{bc-warp}), the $\eta=+1$ case 
has $\psi^{(++)}$ and $\psi^{(--)}$ fermions, and 
the $\eta=-1$ case $\psi^{(+-)}$ and $\psi^{(-+)}$. The fields which
obey the Neumann boundary conditions at the UV 
brane, i.e.\ $\psi^{(++)}$ and $\psi^{(+-)}$ have non-vanishing
boundary values on the UV brane and act as sources of corresponding
CFT operators. In addition, due to the presence of the UV brane, the
source fields obtain kinetic terms at quantum level and become
dynamical. On the other hand, the Dirichlet boundary condition at the
UV brane implies that $\psi^{(--)}$ and $\psi^{(-+)}$ do not take
nonzero values on the UV brane and therefore they are not fundamental
degrees of freedom in the UV regime of CFT\@.

The boundary conditions at the IR brane determine the low-energy
description of CFT\@. The broken CFT generates various bound states.
There are also composite states which have the same quantum number 
of $\psi^{(++)}$ and then only a combination of $\psi^{(++)}$ and these
composite states remains massless. The mixture is determined by the
anomalous dimension of corresponding CFT operator, in other words, a
bulk mass of AdS field. On the other hand, $\psi^{(+-)}$ acquires a
mass term with the corresponding massless CFT bound 
state ${\cal O}^{(+-)}$ and decouples at the IR scale. The 
state ${\cal O}^{(+-)}$ has an opposite quantum number 
to $\psi^{(+-)}$ and corresponds to a dynamical degree of freedom 
of $\psi^{(-+)}$ in the IR region. The field $\psi^{(--)}$ is
understood as a massive bound state.

Now we can see how the anomaly is canceled in the IR. Since the
massless modes at the IR scale are same as those in the effective
four-dimensional theory in the AdS side, the anomaly cancels between
these massless modes. In the UV 
region, $\psi^{(++)}$, $\psi^{(+-)}$, and fields living on the UV
brane contribute to the $G$ gauge anomaly. There is also complicated
CFT contribution to the anomaly and it is then nontrivial how these
anomalies cancel out. However, due to the 't~Hooft anomaly matching
condition, the CFT-induced anomaly can be evaluated by massless
composite states even for strongly-coupled CFT dynamics. The massless
bound states are composed of ${\cal O}^{(+-)}$ and the fields which
are dual to those living only on the IR brane. Since the anomaly 
from $\psi^{(+-)}$ is canceled by that from ${\cal O}^{(+-)}$, the
conditions for anomaly cancellation become identical in the UV and IR
regions of CFT~\cite{CFT-anom}. According to the AdS/CFT
correspondence, the CFT contribution to the $G$ anomaly is mapped to a
Chern-Simons term in the AdS side. Therefore the action in the AdS
side should have a Chern-Simons term whose coefficient is proportional
to the sum of charges of ${\cal O}^{(+-)}$ and fermions localized on
the IR brane. When we act a $G$ gauge transformation, the Chern-Simons
term induces explicit $G$-breaking terms in the divergence of the
current, which are localized on the two boundaries with equal
magnitude and opposite sign. Since in the CFT dual theory the gauge
symmetry $G$ is not broken, the breaking terms should be canceled out
by fermion contributions and $G$ is not broken also in the AdS
side. Thus we recover the localized anomaly in the case that the
effective four-dimensional theory has a vector-like spectrum.

\medskip

We have found that the five-dimensional gauge anomaly on the curved
metric (\ref{warp}) is exactly the same as that in the flat case. It
is independent of the metric factor $a(y)$ and bulk fermion
masses. This result also follows from the theoretical arguments
presented above. We thus conclude that orbifold theories in the
non-factorizable geometries are free from anomalies when
four-dimensional effective theories have anomaly-free spectra.

\section{The Fayet-Iliopoulos term in warped geometry}

In this section we study the Fayet-Iliopoulos (FI) terms in
supersymmetric $U(1)$ gauge theories on the AdS$_5$ geometry. It is
well known that a four-dimensional $U(1)$ theory with charged matter
fields has the one-loop FI term which is quadratically divergent and
proportional to the sum of charges~\cite{FIDterm}. This
radiatively-generated FI term is important for studying the vacuum of
theory and it also has a deep connection to gravitational
anomalies in supergravity theory. Therefore the characteristic
features of FI terms in curved spacetime deserve to be investigated
similarly to the gauge anomalies examined in the previous section. The
one-loop FI terms in flat five-dimensional orbifold theories were
calculated in~\cite{exFID,SSSZ} where they are induced via bulk and
boundary charged fields. We investigate this phenomenon 
in $U(1)$ gauge theories on the warped background. We will
particularly find that FI divergences appear not only on the
boundaries but also in the five-dimensional bulk. An interpretation of
this is that the AdS$_5$ supersymmetry requires different values of
mass parameters for two chiral multiplets contained in one
hypermultiplet.

Once the FI term is generated, it is a non-trivial problem whether one
has a supersymmetric vacuum without breaking other symmetries. For
example, a nonzero four-dimensional FI term necessarily causes
supersymmetry and/or $U(1)$ gauge symmetry breaking. However it has
been shown that, in flat five-dimensional theories, both supersymmetry
and gauge symmetry can survive in spite of the presence of FI terms,
provided that the theory is free from $U(1)$-gravitational mixed
anomaly~\cite{BCCRS,NNO}. The analysis of flat directions shows that
the scalar field in the $U(1)$ vector multiplet develops a vacuum
expectation value due to the FI term. This vacuum expectation value
generates constant bulk masses for charged matter fields and
significantly modifies the wavefunction profiles of KK
modes~\cite{AGW,NNO,MP,wave}, which have important consequences in
low-energy effective theories. We will study a condition for
supersymmetric vacuum and discuss phenomenological implications of the
FI terms in the warped geometry.

\subsection{Superspace action}

We work with supersymmetric quantum electrodynamics in the 
warped (AdS) geometry. Its background metric is given by (\ref{warp})
with the metric factor $a(y)=e^{-k|y|}$. The fifth dimension $y$ has
the two boundaries at $y=0$ and $y=\pi R$ as in the previous
section. To discuss radiative corrections to auxiliary scalar fields,
it is relevant to use the off-shell superspace formalism of
higher-dimensional supersymmetry~\cite{AGW,SFF}. In the superfield
language, the present theory involves two types of five-dimensional
supermultiplets; vector and hyper multiplets. A five-dimensional
off-shell vector multiplet contains a four-dimensional vector 
multiplet $V=(A_\mu,\,\lambda,\,D)$ and a neutral chiral 
multiplet $\chi=(\Sigma+iA_5,\,\lambda',\,F_\chi)$, where $\lambda$
and $\lambda'$ are gauge fermions, and $D$ and $F_\chi$ are auxiliary
scalar fields. An off-shell hypermultiplet consists of
oppositely-charged two chiral multiplets $\Phi=(\phi,\,\psi,\,F_\phi)$
and $\Phi^c=(\phi^c,\,\psi^c,\,F_{\phi^c})$. The five-dimensional
action for AdS$_5$ supermultiplets is given in the $N=1$ superspace
form
\begin{eqnarray}
  S_V &=& \int\! d^4xdy\,\bigg[\int\! d^2\theta\,\frac{1}{4g^2}
  W^\alpha W_\alpha +{\rm h.c.} +\int\! d^2\theta d^2\bar\theta\,
  \frac{e^{-2k|y|}}{g^2}\Big(\partial_yV-\frac{1}{\sqrt{2}}
  (\chi+\chi^\dagger)\Big)^2 \,\bigg], \\[1mm]
  S_H &=& \int\! d^4xdy\,\bigg[ \int\! d^2\theta d^2\bar\theta\,
  e^{-2k|y|} \Big(\Phi^\dagger e^{-qV}\Phi 
  +\Phi^c e^{qV} {\Phi^c}^\dagger \Big) \nonumber \\
  && \hspace{2cm} +\int\! d^2\theta\, e^{-3k|y|} 
  \Phi^c\bigg[\partial_y-\frac{q}{\sqrt{2}}\chi 
  -\Big(\frac{3}{2}-c\Big)k\epsilon(y)\bigg]\Phi +{\rm h.c.}\,\bigg],
\end{eqnarray}
where $q$ and $c$ denote the charge and mass parameter of the chiral
multiplet $\Phi$, and $\epsilon(y)$ is the sign function. The
exponential warp factors have been included explicitly. For the lowest
component of each supermultiplet, these warp factors describe the
metric dependence such as $\sqrt{-g}$ in the AdS$_5$ action, but the
metric factors for other component fields in the AdS$_5$ action are
obtained after the following rescaling
\begin{eqnarray}
\begin{array}{ccccccc}
  \lambda \,\to\, e^{-\frac{3}{2}k|y|}\lambda, &&
  \lambda' \,\to\, e^{-\frac{1}{2}k|y|}\lambda', &&
  D \,\to\, e^{-2k|y|}D, && F_\chi \,\to\, e^{-k|y|}F_\chi, \\[1mm]
  \psi \,\to\, e^{-\frac{1}{2}k|y|}\psi, &&
  \psi^c \,\to\,  e^{-\frac{1}{2}k|y|}\psi^c, &&
  F_\phi \,\to\, e^{-k|y|}F_\phi, &&
  F_{\phi^c} \,\to\, e^{-k|y|}F_{\phi^c}.
\end{array} \label{rescale}
\end{eqnarray}
For later discussion, we mention that after integrating out the
auxiliary fields, the scalar components $\phi$ and $\phi^c$ in a
hypermultiplet have the following five-dimensional masses~\cite{GP}
\begin{eqnarray}  
  m_\phi^2 \,&=& \big(c^2+c-15/4\big)k^2
  +(3-2c)k\big[\delta(y)-\delta(y-\pi R)\big], \\[1mm]
  m_{\phi^c}^2 &=& \big(c^2-c-15/4\big)k^2
  +(3+2c)k\big[\delta(y)-\delta(y-\pi R)\big].
\end{eqnarray}

The boundary conditions imposed on the supermultiplets are similarly
chosen as in Section 2. The vector multiplet $V$ has the Neumann 
boundary conditions at both UV and IR branes and its superpartner
multiplet $\chi$ has the Dirichlet ones because it contains the fifth
component of bulk gauge field. Noting that a boundary condition 
of $\Phi^c$ must be opposite to that of superpartner $\Phi$ for
respecting the $Z_2$ orbifold, we have four different possibilities 
for boundary conditions of hypermultiplets. They are expressed as
($\Phi^{(++)},\,\Phi^{c\,(--)}$), ($\Phi^{(--)},\,\Phi^{c\,(++)}$),
($\Phi^{(+-)},\,\Phi^{c\,(-+)}$), and ($\Phi^{(-+)},\,\Phi^{c\,(+-)}$)
with the notation introduced in the previous section. A hypermultiplet
also carries two other parameters; its $U(1)$ charge $q$ and bulk mass
parameter $c$. It can be seen that the action is irrelevant under the
replacement of ($\Phi^{(+\pm)},\,\Phi^{c\,(-\mp)}$) with ($q,\,c$) by
($\Phi^{(-\mp)},\,\Phi^{c\,(+\pm)}$) with ($-q,\,-c$). Therefore it is
sufficient to show the results only for the two 
cases $\Phi^{(++)}$ and $\Phi^{(+-)}$ with generic values 
of $q$ and $c$. The results for the other parity assignments can be
obtained by simply changing signs of charges and bulk mass parameters.

We also include the action for chiral multiplets confined on the UV
and IR boundaries.
\begin{eqnarray}
  && \hspace*{-8mm} S_{\rm UV} = \int\! d^4xdy\,
  \bigg[ \int\! d^2\theta d^2\bar\theta\, \Phi_{\rm UV}^\dagger 
  e^{-q_{{}_{\rm UV}}\!V}\Phi_{\rm UV} +\!\int\! d^2\theta\, 
  W_{\rm UV}(\Phi,\Phi^c,\Phi_{\rm UV}) +{\rm h.c.} \bigg]\delta(y), 
  \label{bc1} \\
  && \hspace*{-8mm} S_{\rm IR} = \int\! d^4xdy\,
  \bigg[ \int\! d^2\theta d^2\bar\theta\, e^{-2k\pi R}
  \Phi_{\rm IR}^\dagger e^{-q_{{}_{\rm IR}}\!V}\Phi_{\rm IR} 
  +\!\int\! d^2\theta\, 
  e^{-3k\pi R} W_{\rm IR}(\Phi,\Phi^c,\Phi_{\rm IR}) +{\rm h.c.} 
  \bigg]\delta(y-\pi R). \nonumber \\ \label{bc2}
\end{eqnarray}
The warp factors have been taken into account in the IR brane
action. The $Z_2$ boundary conditions break a half of bulk
supersymmetry and thus the boundary actions preserve 
only $N=1$ supersymmetry. The boundary chiral 
multiplets $\Phi_{\rm UV}$ and $\Phi_{\rm IR}$ couple only to bulk
multiplets with Neumann boundary conditions on each brane. We assume
for simplicity that there are no $y$-derivative couplings of $Z_2$-odd
chiral multiplets and no four-dimensional gauge fields on the branes,
though these assumptions are irrelevant to the following discussion.

For abelian gauge theory, the FI term of vector multiplet $V$ can
also be added to the action
\begin{eqnarray}
  S_D &=& \int\! d^4xdy\! \int\! d^2\theta d^2\bar\theta\, 2\xi(y) V 
  \,=\, \int\! d^4xdy\,\xi(y) D.
\end{eqnarray}
We have defined the coefficient $\xi(y)$ into which the metric warp
factor is absorbed. When the fifth dimension is compactified in the
orbifold, $\partial_y V-(\chi+\chi^{\dag})/\sqrt{2}$ becomes a
supergauge invariant combination. The 
additional $(\chi+\chi^{\dag})/\sqrt{2}$ part is superfluous in flat
theories but is not in curved theories. Such a term, however, takes no
part in the following analysis of FI terms.

\subsection{One-loop FI tadpoles}

It is well known in four-dimensional abelian gauge theory that even
if the FI term is set to be zero at the classical level, it is
generated through loop-level divergent tadpole graphs. In the present
five-dimensional case, bulk and boundary scalar fields contribute to
the one-loop FI tadpoles. For bulk scalar fields, the relevant vertex
is found from the action
\begin{equation}
  -\frac{q}{2}e^{-2k|y|}\Big(\phi^\dagger D\phi 
  -\phi^c D{\phi^c}^\dagger \Big),
\end{equation}
which induces a tadpole contribution to the auxiliary field $D$;
\begin{equation}
  \xi(y) \;=\; -\frac{q}{2}e^{-2k|y|}\!\int\!\!\frac{d^4p}{(2\pi)^4}
  \Big[G_p^\phi(y,y)-G_p^{\phi^c}(y,y)\Big].
  \label{tadpole}
\end{equation}
The scalar propagators $G^{\phi,\,\phi^c}_p(y,y')$ on the AdS$_5$ 
background are calculated in the mixed position/momentum space 
where $p$ is the four-dimensional momentum~\cite{GNS,GP2}. Since the
above superspace action already takes account of the metric factors
for scalars, the Green's function evaluated at the coinciding points
in the extra dimension is given by
\begin{eqnarray}
  G^\phi_p(y,y) &=& \frac{i\pi}{2k}e^{4k|y|}
  \bigg[ \widetilde J_{|c+\frac{1}{2}|}\Big(\frac{ip}{k}\Big) 
  H_{|c+\frac{1}{2}|}\Big(\frac{ip}{k}e^{k|y|}\Big) 
  -\widetilde H_{|c+\frac{1}{2}|}\Big(\frac{ip}{k}\Big) 
  J_{|c+\frac{1}{2}|}\Big(\frac{ip}{k}e^{k|y|}\Big) \bigg] 
  \nonumber \\[1mm]
  && \hspace{1.1cm} \times 
  \bigg[ \widetilde J_{|c+\frac{1}{2}|}\Big(\frac{ip}{k}e^{k\pi R}\Big) 
  H_{|c+\frac{1}{2}|}\Big(\frac{ip}{k}e^{k|y|}\Big) 
  -\widetilde H_{|c+\frac{1}{2}|}\Big(\frac{ip}{k}e^{k\pi R}\Big) 
  J_{|c+\frac{1}{2}|}\Big(\frac{ip}{k}e^{k|y|}\Big) \bigg] 
  \nonumber \\[1mm]
  && \hspace{1.1cm} \bigg/ \bigg[ 
  \widetilde J_{|c+\frac{1}{2}|}\Big(\frac{ip}{k}e^{k\pi R}\Big) 
  \widetilde H_{|c+\frac{1}{2}|}\Big(\frac{ip}{k}\Big) 
  -\widetilde H_{|c+\frac{1}{2}|}\Big(\frac{ip}{k}e^{k\pi R}\Big) 
  \widetilde J_{|c+\frac{1}{2}|}\Big(\frac{ip}{k}\Big) \bigg],
\end{eqnarray}
where $H_\alpha$ is the Hankel function of the first kind of 
order $\alpha$ and $J_\alpha$ is the Bessel function. The boundary
condition of $\Phi$ at the UV (IR) brane determines the function forms
of $\widetilde J$ and $\widetilde H$ with the 
argument $ip/k$ ($ipe^{k\pi R}/k$). If $\Phi$ has a Neumann boundary
condition, then $\widetilde J_{|c+1/2|}(z)=\pm zJ_{\pm(c-1/2)}(z)$ for
$\pm(c+1/2)>0$, while if a boundary condition is 
Dirichlet, $\widetilde J_{|c+1/2|}(z)=J_{|c+1/2|}(z)$, and similarly
for $\widetilde H$. The propagator $G_p^{\phi^c}$ of the partner
chiral multiplet of $\Phi$ can be obtained by replacement $c\to-c$ in
all the above expressions.

Given the explicit expressions of Green's functions, we perform the
four-dimensional momentum integral for the FI 
tadpole (\ref{tadpole}), which is at most quadratically divergent. 
We find the leading divergent FI term in the warped geometry, which is
radiatively induced by a bulk hypermultiplet with charge $q$ and mass
parameter $c$ :
\begin{eqnarray}
  \xi(y) &=& \frac{q}{32\pi^2} \Big[\,
  \Lambda^2\big[\delta(y)+\eta\delta(y-\pi R)\big]
  +2ck\Lambda \big[\delta(y)-e^{-k\pi R}\delta(y-\pi R)\big] 
  -ck^2\Lambda e^{-ky}  \nonumber \\
  && \hspace*{5cm} +(kc)^2\ln\Lambda 
  \big[\delta(y)+\eta e^{-2k\pi R}\delta(y-\pi R)\big]\,\Big],
  \label{FI}
\end{eqnarray}
with a sharp UV cutoff $\Lambda$. The parameter $\eta$ takes $+1$ for
the boundary condition $\Phi^{(++)}$ and $-1$ for $\Phi^{(+-)}$. As we
mentioned, the FI term from $\Phi^{(--)}$ [$\Phi^{(-+)}$] can be
derived by changing the signs of $q$ and $c$ in the result 
of $\Phi^{(++)}$ [$\Phi^{(+-)}$]. The net result of FI term is
obtained by summing up all hypermultiplet contributions with different
charges, masses, and boundary conditions. The four-dimensional
boundary fields $\Phi_{\rm UV}$ and $\Phi_{\rm IR}$ also contribute
the divergent FI term. It is easily found from the boundary 
actions (\ref{bc1}) and (\ref{bc2}) that one-loop tadpoles are
calculated in the usual four-dimensional way and given by
\begin{eqnarray}
  \xi(y) &=& \frac{1}{16\pi^2}\Lambda^2 \Big[ q_{{}_{\rm UV}}\delta(y)
  +q_{{}_{\rm IR}}\delta(y-\pi R)\Big].
  \label{FI-boundary}
\end{eqnarray}
The total amount of divergent FI terms is given by the sum 
of (\ref{FI}) and (\ref{FI-boundary}). Here we mention the 
position ($y$) dependence of the cutoff $\Lambda$. A possibility,
suggested by the AdS/CFT correspondence, is that the cutoff varies
with the warp factor. This is implemented by the 
replacement $\Lambda$ with $\Lambda e^{-k|y|}$ in the above formula of
the FI term and also in the analysis below. In particular, after this
replacement the boundary FI terms are consistent with supergravity
analysis~\cite{ACK} and also with the Pauli-Villars regularization.

The FI term due to bulk hypermultiplets (\ref{FI}) shows that the
quadratic divergences are localized at the boundaries. This is
resemblance to the localized gauge anomalies found in the previous
section. In fact, the quadratic divergence is proportional to the sum
of matter charges, which is also proportional to the coefficient of
mixed gravitational anomaly. The relation between FI terms and
gravitational anomalies gives a deep understanding in supergravity
embedding of the result. The embedding would also suggest that
higher-order corrections do not induce additional FI terms. These
issues are beyond the scope of this paper and we leave them to future
investigations. Here we mention a relation between a possible anomaly
cancellation and the structure of $\Lambda^2$ term. In the effective
four-dimensional theory, a mixed gravitational anomaly vanishes if
\begin{eqnarray}
  \sum q_{++}-\sum q_{--}+\sum q_{{}_{\rm UV}}+\sum q_{{}_{\rm IR}}
  &=& 0,
  \label{mixanom}
\end{eqnarray}
where $q_{++}$ and $q_{--}$ are the $U(1)$ charges 
of $\Phi^{(++)}$ and $\Phi^{(--)}$, respectively. This anomaly-free
condition leads to a total $\Lambda^2$ term of the form
\begin{eqnarray}
  \xi_{{}_{\Lambda^2}}(y) &=& \frac{Q}{32\pi^2} 
  \Lambda^2 \big[\delta(y)-\delta(y-\pi R)\big].
  \label{Q}
\end{eqnarray}
The constant $Q$ is given by a relevant sum of charges:
$Q=\sum q_{+-}-\sum q_{-+}+\sum q_{{}_{\rm UV}}-\sum q_{{}_{\rm IR}}$ 
with an obvious notation. One can see a kinship between gravitational
anomalies and FI terms; the anomaly-free spectrum gives rise to 
a common coefficient for the two boundary FI terms. When the
condition (\ref{mixanom}) is satisfied, the theory can be regularized
with a set of Pauli-Villars fields with suitable charges and bulk
masses such that no light mode with wrong statistics is left in the
low-energy theory. Possible candidates for regulator hypermultiplets
must only contain $\Phi^{(+-)}$ [$\Phi^{(-+)}$] with bulk 
masses $c<\frac{1}{2}$ ($c>-\frac{1}{2}$). Other types of fields
contain massless modes in the limit of infinitely large mass
parameters.

The linear FI divergence comes only from hypermultiplets and
represents characteristic properties due to the warped geometry. The
scale suppression by the metric factor exists in the $\Lambda^1$
part. Furthermore, the $\Lambda^1$ divergence is not only confined on
the boundaries but also penetrates into the five-dimensional bulk,
which is quite different from the flat orbifold case where there is no
FI divergence in the bulk. This behavior is also rather different from 
the $\Lambda^2$ term and gauge anomalies. A conceivable understanding
of this fact is based on the mass difference between two scalar fields
in a hypermultiplet. On the AdS background, these two scalars have
different graviphoton charges~\cite{graviphoton} and then have
different masses in the bulk as well as on the branes. If one expands
the integrand of the FI term (\ref{tadpole}) with respect to scalar
masses, the first sub-leading order in terms of $p$ is proportional to
the bulk mass difference. In fact, the bulk $\Lambda^1$ divergence
agrees with this scalar mass difference. The additional exponential
factors come from details of vertices and propagators. We also mention
that in the limit $k\to 0$ with bulk masses $kc$ fixed, the radiative
FI terms in flat theory~\cite{BCCRS} are properly recovered.

\section{Phenomenology of the FI terms}

The $D$-term equation is found from the five-dimensional
supersymmetric action described before,
\begin{eqnarray}
  D &=& -\partial_y(e^{-2k|y|}\Sigma)-g^2\xi(y)
  +\sum\frac{qg^2}{2}e^{-2k|y|}(\phi^\dagger \phi
  -{\phi^c}^\dagger \phi^c) \nonumber \\
  && \hspace{1cm} +\sum\frac{q_{{}_{\rm UV}}g^2}{2}
  \phi_{\rm UV}^\dagger\phi_{\rm UV}\delta(y)
  +\sum\frac{q_{{}_{\rm IR}}g^2}{2}e^{-2k\pi R}
  \phi_{\rm IR}^\dagger\phi_{\rm IR}\delta(y-\pi R),
\end{eqnarray}
where $\Sigma$ is the neutral scalar in the $U(1)$ vector
multiplet. If the theory is free from mixed gravitational anomaly, FI
terms is given by (\ref{Q}). With the implementation of a
position-dependent cutoff, there are supersymmetric vacua in the
presence of charged matter fields. In this paper we introduce
IR-boundary fields with non-vanishing vacuum expectation values so as
to satisfy the $D$-flatness condition. In this vacuum $\Sigma$ also
has an expectation value and drastically changes the bulk field
phenomenology, as will be investigated below. If one introduces
UV-brane chiral multiplets instead of IR-brane ones, $\Lambda$ is
changed to $\Lambda e^{-k\pi R}$ in all the expressions below.

\subsection{Bulk field propagators}

Keeping in mind the relation to anomalies, we find the solution of 
the $D$-flatness condition;
\begin{eqnarray}
  \Sigma(y) &=& -\frac{g^2}{64\pi^2} \epsilon(y)
  \Big( Q\Lambda^2e^{2k|y|}+2Ck\Lambda e^{k|y|} \Big),
  \label{sigma}
\end{eqnarray}
where $Q$ was defined in (\ref{Q}) and $C$ is the sum of $qc$ over all
bulk hypermultiplets. We have simply dropped the logarithmically
divergent term since they are less sensitive to the UV
cutoff.\footnote{Including higher-derivative $\delta$ function terms
would be important for more detailed analysis of KK-mode
wavefunctions. Thank S.~Groot Nibbelink for noticing us this
issue. See also recent work~\cite{SGN}.}
The quadratically divergent part has a solution as we discussed. It is
interesting that the linearly divergent part can also be compatible
with a supersymmetric vacuum thanks to the existence of FI term
spreading into the five-dimensional bulk. Given the vacuum expectation
value of $\Sigma$, hypermultiplets acquire supersymmetry-preserving
mass terms through the bulk superpotential. Since the value 
of $\Sigma$ is no longer constant in the extra dimension, so are the
hypermultiplet masses. Such $y$-dependent masses could drastically
change the KK-mode spectrum and KK phenomenology from that without
FI terms or in flat orbifold theories. Note that boundary chiral
multiplets do not receive any $D$-term contributions from FI
terms. This is understood from the fact that $\Sigma$ obeys the
Dirichlet boundary conditions at the orbifold fixed points.

Let us study the five-dimensional Green's functions of hypermultiplet
scalars. The Green's functions for fermions behave in similar ways as
long as supersymmetry is preserved. After integrating out the
auxiliary components of bulk supermultiplets, we find the propagator 
of $\phi$ (with $U(1)$ charge $q$ and a mass $c$) defined by
\begin{eqnarray}
  \Big[ e^{-2k|y|}\partial_\mu^2 + \partial_y(e^{-4k|y|}\partial_y)
  -e^{-4k|y|}M_\phi^2 \Big]G^\phi(x,x',y,y') &=&
  \delta^{(4)}(x-x')\delta(y-y'),
\end{eqnarray}
where the scalar mass-squared $M_\phi^2$ is
\begin{eqnarray}
  && M_\phi^2 \;=\; \bar m(y)^2+e^{4k|y|}\partial_y
  \big[e^{-4k|y|}\bar m(y)\big],  \label{mphi} \\[1mm]
  && \quad \bar m(y) \,\equiv\, \Big(\frac{3}{2}-c\Big)k\epsilon(y)
  +\frac{q}{2}\langle\Sigma(y)\rangle.
\end{eqnarray}
The mass parameter of $M_{\phi^c}^2$ is given by the 
replacement $q\to -q$ and $c\to -c$. It is now convenient to work with
the Green's function in the mixed position/momentum frame which is
defined by Fourier-transforming with respect to the four-dimensional
momentum
\begin{eqnarray}
  G^\phi(x,x',y,y') &=& 
  \int\!\!\frac{d^4p}{(2\pi)^4} e^{ip(x-x')}G^\phi_p(y,y').
\end{eqnarray}
Introducing the conformal coordinate $z=e^{ky}/k$ 
and $G_p^\phi(z,z')=\exp(\int^z_{z'}\!\frac{\bar m}{kw}dw)
\widetilde G_p^\phi(z,z')$, the above equation becomes
\begin{eqnarray}
  \bigg[\partial_z^2-2\Big(akz+b+\frac{c}{kz}\Big)k\partial_z-p^2\bigg] 
  \widetilde G_p^\phi(z,z') &=& (kz)^3\delta(z-z'),
  \label{gr}
\end{eqnarray}
where we have defined the two dimensionless parameters concerning the
FI contributions; $a=\frac{qg^2}{128\pi^2k}Q\Lambda^2$ 
and $b=\frac{qg^2}{64\pi^2}C\Lambda$. We first consider the most
dominant part of the FI term, i.e.\ $a\neq 0$ and $b=0$. The linear
divergences do not cause meaningful changes in the conclusion we will
show. In this case, the generic solution to the homogeneous equation
is described by the Kummer's hypergeometric function. The solutions in
two regions $z<z'$ and $z>z'$ must satisfy relevant boundary
conditions at the fixed points and also the matching conditions 
at $z=z'$ (the continuity of $\widetilde G$ and a jumping condition
for $\partial_z\widetilde G$). We thus find the general expression of
Green's functions for arbitrary scalar fields
\begin{eqnarray}
  G_p^\phi(y,y') &=& \frac{e^{4ky'}}{2kf(y')}
  \exp\!\Big[\Big(\frac{3}{2}-c\Big)k(y-y')
  -\frac{a}{2}(e^{2ky}-e^{2ky'})\Big] \nonumber \\
  && \qquad \times \bigg[ 
  \widetilde{\cal J}_{c+\frac{1}{2}}\Big(\frac{1}{k}\Big)  
  {\cal H}_{c+\frac{1}{2}}\Big(\frac{e^{ky_{{}_<}}}{k}\Big) 
  -\widetilde{\cal H}_{c+\frac{1}{2}}\Big(\frac{1}{k}\Big)  
  {\cal J}_{c+\frac{1}{2}}\Big(\frac{e^{ky_{{}_<}}}{k}\Big) \bigg] 
  \nonumber \\
  && \qquad \times \bigg[ 
  \widetilde{\cal J}_{c+\frac{1}{2}}\Big(\frac{e^{k\pi R}}{k}\Big) 
  {\cal H}_{c+\frac{1}{2}}\Big(\frac{e^{ky_{{}_>}}}{k}\Big) 
  -\widetilde{\cal H}_{c+\frac{1}{2}}\Big(\frac{e^{k\pi R}}{k}\Big)  
  {\cal J}_{c+\frac{1}{2}}\Big(\frac{e^{ky_{{}_>}}}{k}\Big) \bigg]
  \nonumber \\
  && \qquad \bigg/ \bigg[ 
  \widetilde{\cal J}_{c+\frac{1}{2}}\Big(\frac{e^{k\pi R}}{k}\Big)  
  \widetilde{\cal H}_{c+\frac{1}{2}}\Big(\frac{1}{k}\Big)  
  -\widetilde{\cal H}_{c+\frac{1}{2}}\Big(\frac{e^{k\pi R}}{k}\Big)  
  \widetilde{\cal J}_{c+\frac{1}{2}}\Big(\frac{1}{k}\Big) \bigg],
  \label{green}
\end{eqnarray}
where we have defined $y_{{}_<}$ ($y_{{}_>}$) to be the 
lesser (greater) of $y$ and $y'$. The  
functions ${\cal J}$ and ${\cal H}$ correspond to the two independent
solutions to the homogeneous part of the equation (\ref{gr}) which are
given by the Kummer's function
\begin{eqnarray}
  {\cal J}_\beta(w) \,=\, (k\omega)^{2\beta}
  {}_1F_1\Big(\frac{p^2}{4ak^2}+\beta;\,1+\beta;\,ak^2w^2\Big), &&
  {\cal H}_\beta(w) \,=\, 
  {}_1F_1\Big(\frac{p^2}{4ak^2};\,1-\beta;\,ak^2w^2\Big).\qquad
\end{eqnarray}
In a special case that $\beta=n$ ($n$ : non-positive integer), the two
functions are not independent (corresponding to $J_n$ and $J_{-n}$ in
the flat limit), and ${\cal H}$ should be replaced by some independent
function. We do not consider this special value of parameter but the
analysis below does not lose any generalities. The boundary conditions
on the propagator provide the forms 
of $\widetilde{\cal J}$ and $\widetilde{\cal H}$;
\begin{eqnarray}
  \widetilde{\cal J}_\beta(w) &=& \left\{
  \begin{array}{ccl}
    \frac{\omega}{2}{\cal J}'_\beta(w) && \mbox{for~Neumann}
    \quad \big([\partial_y -\bar m(y)]G_p^\phi=0\big) \\[2mm]
    {\cal J}_\beta(w) && \mbox{for~Dirichlet} \quad (G_p^\phi=0)
  \end{array} \right.
\end{eqnarray}
and similarly for $\widetilde{\cal H}_\beta$. For example, 
if $\phi$ is free on the UV (IR) brane, 
then $\widetilde{\cal J}_\beta(\frac{1}{k})
=\frac{1}{2k}{\cal J}'_\beta(\frac{1}{k})$
$\,\big[\widetilde{\cal J}_\beta(\frac{e^{k\pi R}}{k})
=\frac{e^{k\pi R}}{2k}{\cal J}'_\beta(\frac{e^{k\pi R}}{k})\big]$. With
these functions, $f(y)$ is written as
\begin{eqnarray}
  f(y) &=& \frac{e^{ky}}{2k}\Big[
  {\cal J}'_{c+\frac{1}{2}}\Big(\frac{e^{ky}}{k}\Big)  
  {\cal H}_{c+\frac{1}{2}}\Big(\frac{e^{ky}}{k}\Big) 
  -{\cal H}'_{c+\frac{1}{2}}\Big(\frac{e^{ky}}{k}\Big)  
  {\cal J}_{c+\frac{1}{2}}\Big(\frac{e^{ky}}{k}\Big)\Big].
\end{eqnarray}
In the limit of vanishing FI term ($a\to 0$), the dimensionless
function $f(y)$ behaves as $e^{(2c+1)ky}$.

\subsection{KK spectrum and wavefunction profiles}

\subsubsection{$\Phi^{(++)}$ and $\Phi^{(--)}$ : a massless mode}
\label{++--}

The four-dimensional mass spectrum of chiral supermultiplets are
extracted from pole conditions of the five-dimensional 
propagator (\ref{green}). The conditions crucially depend on the
boundary conditions of chiral multiplets. The general form of pole
conditions becomes
\begin{eqnarray}
  \widetilde{\cal J}_{c+\frac{1}{2}}\Big(\frac{e^{k\pi R}}{k}\Big)  
  \widetilde{\cal H}_{c+\frac{1}{2}}\Big(\frac{1}{k}\Big)  
  -\widetilde{\cal H}_{c+\frac{1}{2}}\Big(\frac{e^{k\pi R}}{k}\Big)  
  \widetilde{\cal J}_{c+\frac{1}{2}}\Big(\frac{1}{k}\Big) &=& 0.  
  \label{pole}
\end{eqnarray}
In this equation, the four-momentum has been replaced by mass
eigenvalue $p^2=-m^2$. For a $\Phi^{(++)}$ multiplet, the equation is
explicitly given by
\begin{eqnarray}
  && \frac{m^2}{4ak^2}\bigg[ e^{(2c-1)k\pi R} 
  {}_1F_1\Big(\frac{-m^2}{4ak^2}+c+\frac{1}{2};\,c+\frac{1}{2};\,
  ae^{2k\pi R}\Big)
  {}_1F_1\Big(\frac{-m^2}{4ak^2}+1;\,\frac{3}{2}-c;\,a\Big) 
  \nonumber \\[1mm]
  && \hspace{10mm} -{}_1F_1\Big(\frac{-m^2}{4ak^2}+c+\frac{1}{2};\,
  c+\frac{1}{2};\,a\Big) 
  {}_1F_1\Big(\frac{-m^2}{4ak^2}+1;\,\frac{3}{2}-c;\,
  ae^{2k\pi R} \Big) \bigg] \;=\; 0.
\end{eqnarray}
It is clearly seen that there is always a massless mode. Since the
present vacuum does not break the low-energy supersymmetry, a
fermionic partner of this scalar zero mode is also massless and they
make up a massless chiral multiplet in four dimensions. On the other
hand, the equation (\ref{pole}) for $\Phi^{(--)}$ turns out to become
\begin{eqnarray}
  && e^{(2c+1)k\pi R} 
  {}_1F_1\Big(\frac{-m^2}{4ak^2}+c+\frac{1}{2};\,c+\frac{3}{2};\,
  ae^{2k\pi R}\Big)
  {}_1F_1\Big(\frac{-m^2}{4ak^2};\,\frac{1}{2}-c;\,a\Big) 
  \nonumber \\[1mm]
  && \hspace{2cm} -{}_1F_1\Big(\frac{-m^2}{4ak^2}+c+\frac{1}{2};\,
  c+\frac{3}{2};\,a\Big) 
  {}_1F_1\Big(\frac{-m^2}{4ak^2};\,\frac{1}{2}-c;\,
  ae^{2k\pi R} \Big) \;=\; 0.
\end{eqnarray}
The eigenvalue equations for a $\Phi^c$ chiral multiplet are obtained
by replacing $a\to -a$ and $c\to -c$. Using a formula for the Kummer's
function ${}_1F_1(\alpha;\beta;\gamma)=
e^\gamma {}_1F_1(\beta-\alpha;\beta;-\gamma)$, it can be checked that
the pole condition of $\Phi^{c\,(--)}$ is identical to that of the
KK-excited modes of $\Phi^{(++)}$. Therefore the KK massive spectrum
of $\Phi^{c\,(--)}$ is paired up with that of $\Phi^{(++)}$, leaving
one four-dimensional massless chiral multiplet in $\Phi^{(++)}$. The
KK-excited modes have a complicated form of mass spectrum depending on
the signs of FI term and bulk mass parameter, but in the case of a
large FI term, the KK mass eigenvalues start from 
around $a^{1/2}$, not suppressed by the exponential warp factor. We
also explicitly checked that a vanishing FI term recovers the mass
eigenvalue equations in the Randall-Sundrum background derived
in~\cite{GP2}.

\medskip

Let us examine the wavefunction profiles of KK chiral multiplets. It
is easily found that all the massive KK modes are localized to the IR
brane for any values of $U(1)$ charges and bulk mass parameters. These
excited modes become rather heavy and may have no significant effects
on the low-energy theory, and therefore we will focus on the massless
eigenstate in $\Phi^{(++)}$. The detailed study of KK wavefunctions,
including the effects of higher-derivative terms [the second line in
(\ref{FI})], will be presented elsewhere. The equation of motion for
the massless mode $\phi_0(x) \chi_0^{{}_{(++)}}(y)$ is given by
\begin{eqnarray}
  \partial_y\Big[e^{-4k|y|}\partial_y(\chi_0^{{}_{(++)}})\Big]
  -e^{-4k|y|}M_\phi^2\chi_0^{{}_{(++)}} &=& 0.
\end{eqnarray}
The mass-squared $M_\phi^2$ was defined by Eq.~(\ref{mphi}). Solving
the equation with the Neumann boundary conditions at the fixed points,
we obtain
\begin{eqnarray}
  \chi_0^{{}_{(++)}}(y) &=& N_0\exp\Big[\Big(\frac{3}{2}-c\Big)k|y|
  -\frac{a}{2}e^{2k|y|}\Big].
  \label{zero++}
\end{eqnarray}
The constant $N_0$ is fixed by the 
normalization $\int\!dy e^{-2k|y|}|\chi_0^{{}_{(++)}}|^2 =1$ so that
the four-dimensional massless mode $\phi_0(x)$ has a canonical kinetic
term. The normalization condition means that $N_0$ contains a 
factor $e^\frac{a}{2}$ for $a>0$ (a 
factor $e^{\frac{a}{2}e^{2k\pi R}}$ for $a<0$), which ensures the flat
limit, $k\to 0$ with $ck$ and $ak$ fixed, reproduces the zero-mode
wavefunction found in the literature. If one considers the boundary 
conditions ($\Phi^{(--)}$,$\,\Phi^{c\,(++)}$) and has a massless mode
in $\Phi^c$, one should reverse the signs of $q$ and $c$ in the solution.

As seen from the above expression, the wavefunction profile of the zero
mode highly depends on $a$. Keeping in mind a warp factor $e^{-2k|y|}$
in the kinetic term of hypermultiplet, we have four different cases
depending on the values of two parameters $a$ and $c$.
\begin{list}{}{%
\setlength{\leftmargin}{3mm}%
\setlength{\parsep}{1mm}%
\setlength{\topsep}{1mm}%
\setlength{\itemsep}{0mm}}
\item \hspace*{-4mm} 
(i) $a>0$, $\,c>\frac{1}{2}$ \ and \ (ii) $a<0$, $\,c<\frac{1}{2}$

In these cases, the zero modes have monotonously varying
wavefunctions. They have peaks at the boundary $y=0$ ($y=\pi R$) for
the case (i) [(ii)].
\item \hspace*{-4mm}
(iii) $a>0$, $\,c<\frac{1}{2}$ \ and \ (iv) $a<0$, $\,c>\frac{1}{2}$

In these cases, the zero-mode wavefunctions have the extreme values
between the two branes, which are the maximum and the minimum for the
cases (iii) and (iv), respectively. It is also found that, in the 
case (iii) [(iv)], the value of the wavefunction at $y=0$ is always
larger (smaller) than that at $y=\pi R$. The peaks of the zero modes
are located at the position
\begin{eqnarray}
  y &=& \frac{1}{2k}\ln\bigg(\frac{\frac{1}{2}-c}{a}\bigg),
\end{eqnarray}
which can take various values between the UV and IR branes, depending
on a relative size of $a$ and $c$. If $|a|$ is very large, the
wavefunction profile is similar to the case (i) or (ii). The situation
is changed for a relatively small value of $|a|$, where the FI-term
coefficient is around the scale on the IR brane. In this case, we find
for the case (iii) that the massless chiral multiplet is peaked at a
middle point between the two branes. On the other hand, for the case
(iv), the massless mode is localized on the both 
boundaries $y=0$ and $y=\pi R$. These characteristic profiles may
provide a novel approach to four-dimensional phenomenology.
\end{list}

\subsubsection{$\Phi^{(+-)}$ and $\Phi^{(-+)}$ : a pair of almost
massless modes}

For the other types of boundary 
conditions $\Phi^{(+-)}$ and $\Phi^{(-+)}$, the pole conditions are
also extracted from the general expression of Green's functions. For 
a $\Phi^{(+-)}$ chiral multiplet, the condition becomes
\begin{eqnarray}
  && \hspace{-1cm} \Big(c^2-\frac{1}{4}\Big)e^{-(2c+1)k\pi R} 
  {}_1F_1\Big(\frac{-m^2}{4ak^2};\,\frac{1}{2}-c;\, ae^{2k\pi R}\Big) 
  {}_1F_1\Big(\frac{-m^2}{4ak^2}+c+\frac{1}{2};\,c+\frac{1}{2};\,a\Big) 
  \nonumber \\[1mm]
  && \hspace{-5mm} -\Big(\frac{m^2}{4k^2}\Big)
  {}_1F_1\Big(\frac{-m^2}{4ak^2}+1;\,\frac{3}{2}-c;\,a\Big) 
  {}_1F_1\Big(\frac{-m^2}{4ak^2}+c+\frac{1}{2};\,c+\frac{3}{2};\,
  ae^{2k\pi R} \Big) \;=\; 0,
  \label{+-}
\end{eqnarray}
and for $\Phi^{(-+)}$, 
\begin{eqnarray}
  && \hspace*{-1cm} \Big(c^2-\frac{1}{4}\Big)e^{(2c-1)k\pi R} 
  {}_1F_1\Big(\frac{-m^2}{4ak^2}+c+\frac{1}{2};\,c+\frac{1}{2};\,
  ae^{2k\pi R} \Big) 
  {}_1F_1\Big(\frac{-m^2}{4ak^2};\,\frac{1}{2}-c;\,a \Big) 
  \nonumber \\[1mm]
  && \hspace*{-5mm} -\Big(\frac{m^2}{4k^2}\Big)
  {}_1F_1\Big(\frac{-m^2}{4ak^2}+c+\frac{1}{2};\,c+\frac{3}{2};\,a\Big) 
  {}_1F_1\Big(\frac{-m^2}{4ak^2}+1;\,\frac{3}{2}-c;\,
  ae^{2k\pi R} \Big) \;=\; 0.
  \label{-+}
\end{eqnarray}
One can easily find from these equations that there are no exactly
massless modes. They have been projected out by the boundary
conditions from the KK spectrum. The pole condition 
for $\Phi^{c\,(+-)}$ [$\Phi^{c\,(-+)}$] is given by 
replacement $a\to-a$ and $c\to-c$ in Eq.~(\ref{+-}) [(\ref{-+})] and
is equivalent to the pole condition 
of $\Phi^{(-+)}$ [$\Phi^{(+-)}$]. Therefore the KK modes 
from $\Phi^{(+-)}$ and $\Phi^{c\,(-+)}$ [$\Phi^{(-+)}$ 
and $\Phi^{c\,(+-)}$] necessarily come in pair. Though almost all of
the KK modes become heavy due to the presence of FI term ($a\neq 0$),
there can be one pair of light chiral multiplets in the low-energy
effective theory. The mass of these light modes is explicitly derived
by expanding the eigenvalue equation~(\ref{+-}) [or (\ref{-+})] with a
large FI term
\begin{eqnarray}
  m^2 &\simeq& 4a^2k^2 \exp\!\Big[(1-2c)k\pi R-a(e^{2k\pi R}-1)
  +\frac{\pi Rm^2}{ak}\Big].
\end{eqnarray}
If $a>0$, the last factor in the right-handed side can be dropped and
we have
\begin{eqnarray}
  m &\simeq& 2ak\exp\!\Big[\Big(\frac{1}{2}-c\Big)k\pi R
  -\frac{a}{2}(e^{2k\pi R}-1)\Big].
  \label{zero+-}
\end{eqnarray}
Our derivation shows that such light modes appear in two types of
hypermultiplets; ($\Phi^{(+-)}$,$\,\Phi^{c\,(-+)}$) with a positive
charge $a>0$ and ($\Phi^{(-+)}$,$\,\Phi^{c\,(+-)}$) with a negative
charge $a<0$. The wavefunctions of the light chiral multiplets are
shown to have the following form
\begin{eqnarray}
   \chi_l^{{}_{(+-)}} &\simeq& N_l\exp\Big[\Big(\frac{3}{2}-c\Big)k|y|
  -\frac{a}{2}e^{2k|y|}\Big], \\
  \chi_l^{{}_{c\,(-+)}} \! &\simeq& N^c_l\exp\Big[
  \Big(\frac{3}{2}+c\Big)k|y| +\frac{a}{2}e^{2k|y|}\Big],
\end{eqnarray}
for a positive $a$. The normalization constants $N_l$ and $N_l^c$ are
determined by $\int\!dy e^{-2k|y|}|\chi_l^{{}_{(+-)}}|^2=1$, etc. Note
that the wavefunctions of $\chi_l^{{}_{(-+)}}$ 
and $\chi_l^{{}_{c\,(+-)}}$ with a negative $a$ are exactly the same
as the above, but in the expression of their mass eigenvalue the signs
of $a$ and $c$ must be changed 
from (\ref{zero+-}). The $\chi_l^{{}_{(+-)}}$ wavefunction is similar
to the massless mode $\chi_0^{{}_{(++)}}$ derived 
in (\ref{zero++}) where $U(1)$ charge parameter $a$ was free. This
similarity is because the mass (\ref{zero+-}) is negligibly small 
and $\chi_l^{{}_{(+-)}}$ and $\chi_0^{{}_{(++)}}$ satisfy almost the
same equation of motion. As seen from the above expressions, the
eigenfunction is suppressed at the boundary where the Dirichlet
boundary condition is imposed and has a large value at the opposite
boundary.

We have found, for any values of charges and bulk masses, there are
two light chiral multiplets with a tiny mass between these two. For a
large FI term, $\chi_l^{{}_{(+-)}}$ and $\chi_l^{{}_{c\,(-+)}}$ are
strongly localized, with little overlap, onto the UV and IR branes,
respectively. The resultant spectrum 
for ($\Phi^{(+-)}$,$\,\Phi^{c\,(-+)}$) 
or ($\Phi^{(-+)}$,$\,\Phi^{c\,(+-)}$) is like the supersymmetric
quantum chromodynamics though we have imposed the chiral $Z_2$
projection on the spectrum.

If one takes the flat background 
limit ($k\to 0$ with $ck\equiv\bar c$ and $ak\equiv\bar a$ fixed), the
tiny mass eigenvalue (\ref{zero+-}) becomes
\begin{eqnarray}
  m &\simeq& 2\bar a e^{-(\bar a+\bar c)\pi R}.
\end{eqnarray}
This corresponds to an exponentially-suppressed KK mass found in
Ref.~\cite{MP} for a vanishing bulk mass parameter ($\bar c=0$).

\subsection{Effective theory and physical implications}

In the previous section, we found three types of (almost) massless
chiral multiplets in the low-energy effective 
theory. $\Phi^{(++)}$ has a massless mode independent of $a$, but its
wavefunction highly depends on $a$. With a 
positive (negative) $a$ parameter, we have an almost massless
vector-like modes in $\Phi^{(+-)}$ and $\Phi^{c\,(-+)}$ ($\Phi^{(-+)}$
and $\Phi^{c\,(+-)}$). The 
multiplets $\Phi^{(++)}$, $\Phi^{(+-)}$ with $a>0$ 
and $\Phi^{c\,(+-)}$ with $a<0$ are localized at the UV brane and the
others are at the IR brane. As for a $U(1)$ neutral multiplet 
with $(++)$ boundary condition, it gives a massless chiral multiplet
and the wavefunction is controlled by the bulk mass parameter $c$. In
addition, there is a four-dimensional massless vector 
multiplet $V_0$ whose wavefunction is constant along the extra
dimension. We schematically describe low-energy effective theories for
these light modes as well as four-dimensional boundary chiral
multiplets with the following action
\begin{eqnarray}
  S^{\rm eff}_{\rm vector} &=& \int\!d^4x\!\int\!d^2\theta\,
  \frac{1}{4g_4^2}W_0^\alpha W_{0\alpha} +{\rm h.c.}, \\
  S^{\rm eff}_{\rm chiral} &=& 
  \int\!d^4x\!\int\!d^2\theta d^2\bar\theta\,\bigg[
  \Phi_0^\dagger e^{-q_0V_0}\Phi_0+ \Phi_l^\dagger e^{-q_lV_0}\Phi_l+
  \Phi_l^c e^{q_lV_0}\Phi_l^{c\,\dagger} \nonumber \\
  && \hspace*{1cm} +K_{\rm UV}^{\rm eff} \Big(\Phi_{\rm UV},\Phi_0^{a>0},
  \chi_0^{{}_{(++)}}(0)\Phi_0^{a<0},\Phi_l\Big) \nonumber \\
  && \hspace*{1cm} +e^{-2k\pi R}K_{\rm IR}^{\rm eff} \Big(\Phi_{\rm IR},
  \chi_0^{{}_{(++)}}(\pi R)\Phi_0^{a>0},
  \Phi_0^{a<0},\Phi_l^c\Big) \bigg] \nonumber \\
  && \hspace*{0.5cm} +\int\!d^4x\!\int\!d^2\theta\bigg[
  W_{\rm UV}^{\rm eff} \Big(\Phi_{\rm UV},\Phi_0^{a>0},
  \chi_0^{{}_{(++)}}(0)\Phi_0^{a<0},\Phi_l\Big) \nonumber \\
   && \hspace*{1cm} 
  +e^{-3k\pi R}W_{\rm IR}^{\rm eff} \Big(\Phi_{\rm IR},
  \chi_0^{{}_{(++)}}(\pi R)\Phi_0^{a>0},
  \Phi_0^{a<0},\Phi_l^c\Big) \bigg] +{\rm h.c.},
  \label{4Dchiral}
\end{eqnarray}
where $g_4$ is the gauge coupling defined by $1/g_4^2=\pi R/g^2$. We
have not explicitly written down $O(1)$ factors of wavefunctions for
notational simplicity. The four-dimensional field $\Phi_0(x)$ denotes
a light degree of freedom 
from $\Phi^{(++)}$, and $\Phi_l(x)$ and $\Phi_l^c(x)$ come from the
multiplets with boundary conditions $(+-)$ and $(-+)$ and are
localized at the UV and IR branes, respectively. In the boundary
K\"ahler potentials and superpotentials, only the multiplets with
Neumann boundary conditions appear with relevant strengths on each
brane. The extension to multi flavors of hypermultiplets is
straightforward. The K\"ahler 
potentials $K_{\rm UV}^{\rm eff}$ and $K_{\rm IR}^{\rm eff}$ contain
the kinetic terms for boundary 
multiplets $\Phi_{\rm UV}$ and $\Phi_{\rm IR}$, and also include
possible higher-dimensional operators among bulk and brane
superfields. The 
superpotentials $W_{\rm UV}$ and $W_{\rm IR}$ preserve only half of
bulk supersymmetry and they can involve mass, Yukawa, and other terms
allowed in four-dimensional supersymmetric theory. The superspace form
is relevant to study various phenomenology in the low-energy effective
theory preserving supersymmetry.

The above effective action shows that the vector-like 
multiplets $\Phi_l$ and $\Phi_l^c$ behave in a similar way to boundary
chiral multiplets as long as their charge parameters $a$'s are not so
small. On the other hand, $\Phi_0$ multiplets appear with the factors
of wavefunctions evaluated at the boundaries. The wavefunction factors
appeared in (\ref{4Dchiral}) are
\begin{eqnarray}
  \chi_0^{{}_{(++)}}(0) \,\, &\simeq& \sqrt{\frac{2ke^{-a}}{\ln(-a)}}\,
  \lambda^{\frac{1}{2}(c-\frac{3}{2})}\,
  e^{+\frac{a}{2\lambda}} \qquad (a<0), \label{sup} \\
  \chi_0^{{}_{(++)}}(\pi R) &\simeq& \sqrt{\frac{2k e^a}{\ln a}}\,
  \lambda^{\frac{1}{2}(c-\frac{3}{2})}\,
  e^{-\frac{a}{2\lambda}} \qquad (a>0). \label{sup2}
\end{eqnarray}
We have defined a small parameter $\lambda=e^{-2k\pi R}$. These
wavefunction factors give sources to generate two different sizes of
hierarchies, which are given by powers and exponentials of the ratio
between the UV and IR scales. Taking various values of charges and
bulk masses, we obtain hierarchical quantities in the low-energy
effective theory. This provides a new tool for low-energy
phenomenology.

\smallskip

As the first example, we discuss a large hierarchy among the Yukawa
couplings of quarks and leptons. While the masses of quarks and
charged leptons are in the vicinity of the electroweak scale, the
recent experimental observations indicate that neutrinos have tiny
masses of $O({\rm eV})$, which are $10^{-(11-14)}$ times smaller than
the electroweak scale. Our aim in the current example is to explain
such a huge hierarchy between the masses of neutrinos and other
fermions by exploiting the suppression factors 
of (\ref{sup}) and (\ref{sup2}). As an illustration, we assume that
the electroweak Higgs fields and Yukawa terms are confined on the IR
brane. In addition, the quarks and charged leptons come from the zero
modes of bulk hypermultiplets with $(++)$ boundary conditions, mass
parameters $c_i$, and vanishing $U(1)$ charges.\footnote{In fact, it
is not necessarily to place the entire matter multiplets in the
five-dimensional bulk. A simple way to produce the right fermion
masses is to only put a set of 10 representations of $SU(5)$ in
additional spatial dimensions~\cite{10rep}.}
In this case, the bulk masses $c_i$ determine the wavefunction factors
of charged fermions at the IR brane $\chi_0^{++}(\pi R)$ and hence
generate Yukawa hierarchies between the three generations. This is
archived when the radius modulus is stabilized so 
that $\lambda\sim 10^{-(1-2)}$. We also introduce right-handed
neutrino multiplets with positive $a$ parameters. (The $U(1)$ gauge
invariance requires at least one extra standard-model singlet (or a
doublet Higgs) with a negative $a$ parameter.) The 
non-vanishing $a$ induces a large suppression of neutrino
wavefunctions at the IR brane and the tiny neutrino Yukawa couplings
follow. It is found from (\ref{sup2}) that the suppression is roughly
given by the exponential factor $e^{-a/2\lambda}$ which can give a
correct order of magnitude for neutrino masses with an $O(1)$ value 
of $a$. We also note that a mild hierarchy between the three neutrino
masses is realized by choosing different neutrino bulk 
masses $c_{\nu_i}$.

Another application is concerned with the Planck/weak mass
hierarchy. We now want to show a simultaneous realization both of
Yukawa and Planck/weak mass hierarchies, where the former is generated
by the difference of wavefunctions caused by bulk matter masses
and therefore the parameter $\lambda$ is set to be about a unit of
Yukawa hierarchy. The Plank/weak mass hierarchy is then realized 
by $U(1)$ charge parameter $a$. As a primitive example, let us
consider that five-dimensional boundary superpotentials include the
following terms for Higgs fields
\begin{eqnarray}
  W_{\rm UV} \;=\; \frac{f_S}{\sqrt{M}}
  S(x,y)H_u(x)H_d(x), && W_{\rm IR} \;=\; M^{3/2}S(x,y),
  \label{potential}
\end{eqnarray}
where $f_S$ is a dimensionless coupling and $M$ denotes the
fundamental scale of five-dimensional theory, which is not much
different from the Plank scale. We have assumed that the doublet
Higgses $H_u$ and $H_d$ are confined on the UV brane and the
standard-model singlet $S$ comes from a bulk hypermultiplet 
with $(++)$ boundary condition. Now $S$ and $H_d$ have opposite charge
parameters $a$ and $-a$, respectively ($a>0$) so that the
dimension-four term $W_{\rm UV}$ is gauge 
invariant.\footnote{The $U(1)$ symmetry is softly broken 
by $W_{\rm IR}$. A potential alternative is $U(1)_R$ symmetry under
which $S$ has charge $+2$ and both $W_{\mbox{\tiny UV, IR}}$ are
invariant. In this case, we are necessarily lead to supergravity theory.}
One-loop chiral anomalies can be canceled, e.g.\ by adding appropriate
charged multiplets to boundary theories. Given the charge 
assignment, $S$ contains a zero mode localized at the UV brane. It is
found from (\ref{4Dchiral}) that the supersymmetric vacuum in four
dimensions satisfies
\begin{eqnarray}
  \langle H_u\rangle\langle H_d\rangle &=& 
  \frac{-1}{f_S}\bigg(\frac{\chi_s(\pi R)}{\chi_s(0)}\bigg)
  e^{-3k\pi R}  M^2,
\end{eqnarray}
where $\chi_s$ denotes the wavefunction of the zero mode of $S$ whose
form is read from (\ref{sup2}). Such a wavefunction factor of $S$
appears because the massless mode needs to propagate the bulk space to
communicate the coupling at the IR brane to the UV
brane. Supersymmetry is not broken though we have a linear
superpotential in $W_{\rm IR}$. This is understood by solving 
the $F$-flat conditions explicitly and also from the four-dimensional
effective action. For example, a vanishing bulk mass parameter of the
$S$ field leads to the vacuum expectation values of the doublet Higgses
\begin{eqnarray}
  \langle H_u\rangle, \, \langle H_d\rangle &\simeq& 
  \lambda^{3/8}e^{-\frac{a}{4\lambda}}M,
  \label{hierarchy}
\end{eqnarray}
The parameter $\lambda$ defined before gives a unit of Yukawa
hierarchy. Eq.~(\ref{hierarchy}) thus shows that the Planck/weak mass
hierarchy is obtained by choosing $O(1)$ $a$ parameter.

The Yukawa couplings of quarks and leptons come from the UV boundary
term
\begin{eqnarray}
  W_{\rm UV} &=& \frac{1}{M}f^{ij}_uQ_iu_jH_u 
  +\frac{1}{M}f^{ij}_dQ_id_jH_d +\frac{1}{M}f^{ij}_lL_ie_jH_d
\end{eqnarray}
with the standard notation. All the matter multiplets are assumed to
reside in the bulk and the couplings $f_{u,d,l}$ are 
dimensionless $O(1)$ parameters. For these terms to be gauge
invariant, the $U(1)$ charges of quark and lepton multiplets 
are $a_Q=a_u=a_e=0$ and $a_d=a_L=+a$~($>0$). For simplicity we 
assume $U(1)_Y$ of the standard model gauge group does not have FI
term. The zero modes of $d$ and $L$ are then localized at the UV brane
and do not provide any suppressions of Yukawa couplings $f_d$ 
and $f_l$. We take the mass parameters of $U(1)$-neutral 
multiplets $c_{Q,u,e}<\frac{1}{2}$ so that they are peaked at the IR
brane (see the localization property discussed in
Section~\ref{++--}). As a result, the zero-modes quarks and leptons
are found to have the following Yukawa couplings
\begin{eqnarray}
  (f^{ij}_u)_0 \;\simeq\; 
  \frac{\lambda^{1-c_{{}_{Q_i}}-c_{u_j}}}{\pi MR}, \qquad 
  (f^{ij}_d)_0 \;\simeq\; 
  \frac{\lambda^{\frac{1}{2}-c_{{}_{Q_i}}}}{\pi MR}, \qquad 
  (f^{ij}_l)_0 \;\simeq\; \frac{\lambda^{\frac{1}{2}-c_{e_j}}}{\pi MR}.
  \label{lowYukawa}
\end{eqnarray}
Several interesting results follow from this expression. First, the
hierarchy of up-quark masses is generally larger than that of down
quarks due to the suppression effect of the right-handed up quarks, 
while not disturbing the smallness of quark mixing angles. On the
other hand, the hierarchy of lepton Yukawa couplings are controlled
only by the right-handed electrons. This implies that the left-handed
leptons can largely mix with each other, which is indeed suggested by
the recent experimental results for neutrino physics. If one takes the
bulk masses $c_Q=c_u=c_e$ motivated by grand unification, the
effective Yukawa couplings (\ref{lowYukawa}) predict the mass
eigenvalues and mixing angles roughly consistent with the present
experimental data. We comment that $W_{\rm UV}$ in (\ref{potential})
could also lead to a natural $\mu$-term generation by introducing
additional terms of $S$ on the IR brane.

The mechanism proposed above is different from other approaches to the
hierarchy problems with the AdS warp factor. In
Refs.~\cite{GP2,RS-Yukawa}, the Higgs 
fields (and Yukawa couplings) are confined on the TeV brane to have
the electroweak scale from the warp factor \`a la
Randall-Sundrum~\cite{RS}. The matter multiplets have suitable bulk
masses $c_i$ and are localized at the Planck brane. The situation
generates Yukawa hierarchy in unit of a ratio between the Planck/weak
mass hierarchy, 
namely, $f_{ij}\sim (M_{\rm weak}/M_{\rm Pl})^{c_i+c_j-1}$, and
therefore needs some fine-tuning of mass parameters $c_i$ for
realistic fermion masses and mixing.

\section{Conclusions and discussions}

In this paper we have discussed chiral gauge anomalies in curved
spacetime and also the structure of Fayet-Iliopoulos terms in warped
supersymmetric theories. The chiral anomalies have been found to
appear at the orbifold fixed points and are localized in the exactly
same fashion as in flat theories. This result has the theoretical
supports from a low-energy effective theory viewpoint and also
from the AdS/CFT correspondence. The gauge anomaly is canceled by an
appropriate Chern-Simons term if the four-dimensional effective theory
is free from gauge anomaly.

Unlike the gauge anomaly, the FI terms behave differently from the
flat theory. In the warped geometry, the FI divergence are generated
not only on the boundaries but also in the whole bulk. The effect of
the FI term is to generate supersymmetric masses for charged bulk
multiplets which depend on the metric factor. Typical KK masses for
charged particles are around the scale of FI coefficients. The
wavefunction profiles of bulk multiplets have also been examined for
various types of boundary conditions on the branes. For example, the
massless mode in $\Phi^{(++)}$ with even-even parity is strongly
localized onto the UV or IR brane, depending on its $U(1)$ charge and
bulk mass parameter. We have shown by explicit constructions that this
localization behavior has interesting applications to the problems in
particle physics, e.g., Yukawa and Planck/weak mass hierarchies.

It may be interesting to study the localized anomaly and FI terms via
the four-dimensional theory space approach~\cite{decon}. The field
theory in the AdS$_5$ spacetime can also be formulated along this
line~\cite{RSdecon} and it should be possible to examine gauge
anomalies as in~\cite{decon-anom}. The structure of FI terms is also
understood in a similar way. 

In this paper we have not considered supersymmetry breaking and
gravity. The localized FI terms and resultant modified wavefunctions
significantly affect supersymmetry breaking in the low-energy
theory. Several ways to break supersymmetry are expected in the AdS
background, e.g., orbifolding, brane-localized interactions, the
radius modulus $F$ term. Each mechanism could lead to distinguishable
sparticle spectrum in the low-energy effective theory. Combined with
other phenomenological issues, realistic model construction along this
line may deserve to be investigated. Proceeding to supergravity
theory, we could study more clearly the relations between FI terms
and gravitational anomalies. The invariance under local
supersymmetry might suggest a possible form of the coefficient of FI
term. To examine gravitational back-reactions due to FI terms may also
be an interesting issue.

\vspace*{5mm}
\subsection*{Acknowledgments}

The authors wish to thank B.~Holdom, S.~Groot Nibbelink, E.~Poppitz,
N.~Uekusa and N.~Yamashita for valuable discussions, and are also 
grateful to the organizers of the workshop ``Brane-world and
Supersymmetry'' held at University of British Columbia in Canada 
(July, 2002) where a part of this work was carried out. 
This work is supported in part by the Natural Sciences and Engineering
Research Council of Canada.


\end{document}